\documentclass[a4paper]{jpconf}
\usepackage{iopams} 
\usepackage{psfrag, tabls, dcolumn, bm, color, subfig}
\usepackage[pdftex]{graphicx}           
\usepackage{hyphenat}
\usepackage[english]{babel}
\usepackage{array}

\newcommand{\bgn}{\begin{equation}}
\newcommand{\egn}{\end{equation}}
\newcommand{\bgna}{\begin{eqnarray}}
\newcommand{\egna}{\end{eqnarray}}
\newcommand{\bsplit}{\begin{split}}
\newcommand{\esplit}{\end{split}}

\begin{document}

\title{Real-time switching between multiple steady-states in quantum transport }

\author{A.-M. Uimonen$^1$, E. Khosravi$^{2,5,6}$, G. Stefanucci$^{3,5}$, S. Kurth$^{4,5}$, R. van Leeuwen$^{1,5}$, E.K.U. Gross$^{2,5,6}$}

\address{$^1$Department of Physics, Nanoscience Center, FIN 40014, University of Jyv\"askyl\"a,
Jyv\"askyl\"a, Finland}
\address{$^2$Institut f\"ur Theoretische Physik, Freie Universit\"at Berlin, Arnimallee 14, 14195, Berlin, Germany}
\address{$^3$Dipartimento di Fisica, Universit\`a di Roma Tor Vergata, Via della Ricerca Scientifica 1, I-00133 Rome, Italy}
\address{$^4$Nano-Bio Spectroscopy Group, Dpto. de F\'{i}sica de Materiales, 
Universidad del Pa\'{i}s Vasco UPV/EHU, Centro Mixto CSIC-UPV/EHU, 
Av. Tolosa 72, E-20018 San Sebasti\'{a}n, Spain \\
IKERBASQUE, Basque Foundation for Science, E-48011 Bilbao, Spain}
\address{$^5$European Theoretical Spectroscopy Facility (ETSF)}
\address{$^6$Max Planck Institute of Microstructure Physics, Weinberg 2, 06120, Halle, Germany}

\ead{anna-maija.uimonen@jyu.fi}

\begin{abstract}
We study transport through an interacting model system consisting of a central correlated site
coupled to finite bandwidth tight-binding leads, which are considered as effectively noninteracting. 
Its nonequilibrium properties are determined by real-time propagation of the Kadanoff-Baym equations
after applying a bias voltage to the system.
The electronic interactions on the central site
are incorporated by means of self-energy approximations at Hartree-Fock, second Born and GW level. 
We investigate the conditions under which multiple steady-state solutions occur within 
different self-energy approximations, and analyze in detail the nature of these
states from an analysis of their spectral functions. At the Hartree-Fock level at least two stable steady-state
solutions with different densities and currents can be found. 
By applying a gate voltage-pulse at a given time we are able to switch between these solutions. 
With the same parameters we find only one steady-state solution when the self-consistent second Born and
GW approximations are considered. We therefore conclude that treatment of many-body interactions 
beyond mean-field can destroy bistability and lead to qualitatively different results as compared those at
mean-field level.
\end{abstract}


\section{Introduction}

The experimental observation \cite{Goldman_PRL_87,sheard__APL_88} of a hysteresis 
loop in the I/V characteristic of double-barrier resonant tunneling structures prompted 
intense theoretical activities to gain a microscopic understanding of this phenomenon.
Several authors have been able to reproduce the hysteresis behavior by treating 
the Coulomb interaction at a mean field level \cite{sf.1990,fj.1992,pf.1993,kluksdahl}. Self-consistent calculations 
have revealed the presence of bistable solutions, one 
of the solution being characterized by a considerable accumulation of charge in 
the potential well. Subsequent experimental work on double-barrier resonant tunneling 
diodes has, however, demonstrated that hysteresis loops do not always occur \cite{eetal.1988}. 
The suppression of the intrinsic bistability has been attributed to exchange-correlation 
effects \cite{zwlhcg.1994}.

With the advent of molecular electronics \cite{DiVentra_book} the study of intrinsic bistability 
in nanoscale devices has gained attention due to the possibility of developing, 
for example, molecular diodes. So far, most of the work has focused on the steady-state I/V curve 
of molecular junctions attached to metallic leads. Up to date calculations are performed 
within the one-particle scheme of time-independent density functional theory (DFT).
At the Hartree level bistability was reported by Negre {\em et. al} 
for a double quantum dot structure \cite{Negre_CPL_08}.  

The mechanism of bistability and the calculation of
switching times between two different states are mostly unexplored and the question
how correlations affect the bistability has received very scarce attention. The purpose of the present 
exploratory paper is twofold: to extend  the analysis to the time-domain and to study the role of 
memory effects in a bistable interacting resonant level model (IRLM).

Two complementary theoretical approaches will be used for calculating the time-dependent 
current and density, namely Time-Dependent (TD) DFT and Many-Body Perturbation Theory (MBPT). 
TDDFT \cite{Gross_book} provide an exact framework to account for correlation effects both
in the leads and the central region \cite{Gianluca_PRB_04}. Within TDDFT 
the basic quantities that are propagated in time are the one-particle orbitals which depend on 
only one time variable. This property renders the implementation computationally 
favorable \cite{ksarg.2005}. Most approximations to the TDDFT potential, however,
do not include memory effects  
and the exchange-correlation part is approximated by local or semi-local functionals of the density.
The lack of more sophisticated approximations represents, at present, a major obstacle for an accurate 
first principle description of TD quantum transport through interacting regions.
MBPT has the advantage over TDDFT of allowing for a systematic inclusion of relevant 
physical processes through a selection of Feynman diagrams. Thus, MBPT provides an important 
tool to proceed beyond the commonly used adiabatic approximations and to quantify the importance 
of memory effects through advanced approximations to the self-energy. 
We recently proposed \cite{Petri, petri_epl_08} a time dependent MBPT formulation of quantum
transport, based on the real-time propagation of the Green function \cite{baym_book, danielewicz1, dvl.2007}
for open and interacting systems. First the Dyson equation for the connected system is solved to self-consistency 
on the imaginary axis. After that the Green function is propagated with the Kadanoff-Baym equations 
using different level of conserving approximations. As the Green function 
depends on two time variables the implementation of the MBPT scheme is more 
demanding than that of the TDDFT scheme. We expect that the interplay between 
MBPT and TDDFT will be essential to develop accurate approximation for systems 
more complex than the one studied here.

\section{Interacting Resonant Level Model}

We study an Anderson-type of system \cite{anderson_PR_61} where the impurity 
is an interacting site coupled to the infinite one dimensional tight-binding leads 
of finite band width. The Hamiltonian of the system reads
\begin{eqnarray}
\label{eq:hamiltonian}
 \hat{H}(t)&=& \sum_{\sigma}\big[\varepsilon_{0}+V_g(t)\big]\hat{d}_{0\,\sigma}^{\dagger}\hat{d}_{0,\sigma}
+ \frac{1}{2} \sum_{\sigma,\sigma'} \mathcal{U}\, \hat{d}^\dagger_{0,\sigma} \hat{d}^\dagger_{0,\sigma'} \hat{d}_{0,\sigma'} \hat{d}_{0,\sigma} + \sum_{i,\alpha,\sigma} \big[ a+ U_{\alpha}(t) \big]\hat{c}_{i\alpha,\sigma}^{\dagger}\hat{c}_{i\alpha,\sigma}
 \nonumber \\
 &+&  \sum_{i,\alpha,\sigma}\big[ b \hat{c}_{i\alpha,\sigma}^{\dagger}\hat{c}_{i+1\alpha,\sigma}+h.c. \big]
+ \sum_{\alpha,\sigma}[V_{0,1\alpha}\hat{d}_{0,\sigma}^{\dagger}\hat{c}_{1\alpha,\sigma}+h.c.] ,
\end{eqnarray}    
where $i$ denotes the site indices and $\sigma$ is the spin index, $\varepsilon_{0}$ is the on-site energy 
of the localized site, $\mathcal{U}$ is the strength of the two-particle interaction on the central site, $b$ is the 
hopping parameter between lead sites, $U_{\alpha}(t)$ is time-dependent bias
voltage in the leads $(\alpha=L/R)$, $a$ is the on-site parameter in the leads and $V_{0,1\alpha}$ 
denotes coupling between the lead and the localized site. 
The fermionic creation- and annihilation operators in the leads $\alpha$ are denoted as $\hat{c}^{\dagger}$, $\hat{c}$ 
whereas for the localized site they are denoted as $\hat{d}^{\dagger}$, $\hat{d}$. The quantity $V_g(t)$ 
denotes a time-dependent gate voltage.

The main reason for studying
the IRLM is that for this system the multiple steady-state solutions 
are easily found from a fixed-point equation for the density on the localized site.
The IRLM is the simplest system in which bistability occurs and hence allows for a clear interpretation 
of its multiple steady-state solutions. 
We study the system at a
finite temperature and under a finite bias, such that we are out of the Kondo regime
in which the IRLM is often used.

\begin{figure}[htbp]
  \centering
  \subfloat[]{\label{fig:system}\includegraphics[width=0.45\textwidth]{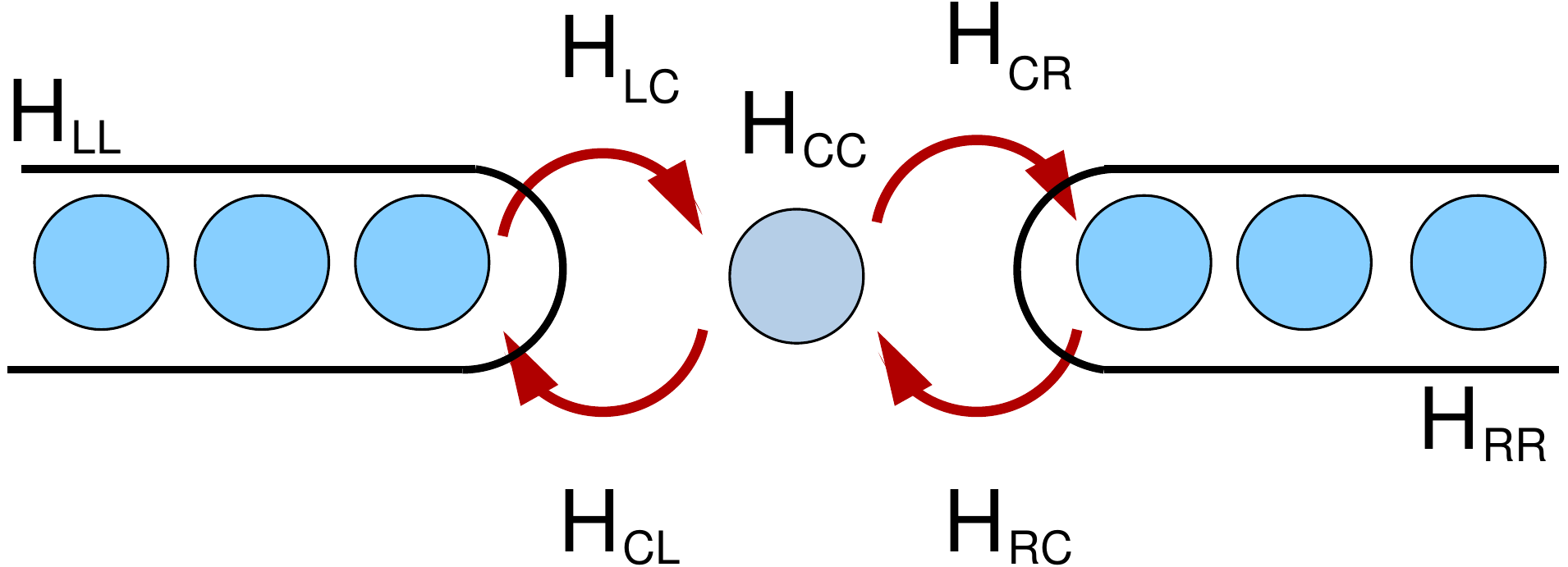}}                
  \subfloat[]{\label{fig:contour}\includegraphics[width=0.53\textwidth]{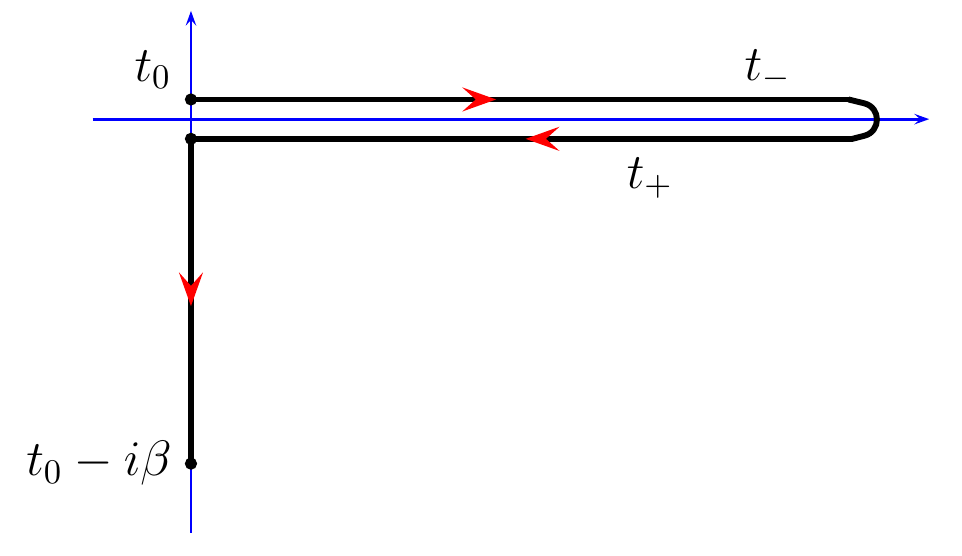}}
  \caption{a) A schematic representation of the studied system. The correlated central region (C) is coupled to 
  infinite one dimensional left (L) and right (R) tight-binding leads via a coupling Hamiltonian. 
  b) The Keldysh contour $c$. Times on the lower branch $t_+$ are later than times on the upper
  branch $t_-$. The imaginary track extends up to the inverse temperature $\beta$.}
  \label{fig:figures(1)}
\end{figure}

\section{Theoretical approach}
\subsection{Kadanoff-Baym equations}

We study the nonequilibrium properties of the IRLM by means of time-propagation of the Kadanoff-Baym 
equations  for the nonequilibrium Green function~\cite{robert_06}.
We assume the system to be contacted and in equilibrium at a chemical potential $\mu$ and at inverse temperature
$\beta$ before the time $t=t_0$. For $t>t_0$ 
the system is driven out of equilibrium by an external bias and we aim to study the time-evolution of the 
electron density and current. We give here a brief description of the approach described in more detail
in Ref.~\cite{Petri}.

The Keldysh Green function is defined as the expectation value of the contour ordered product~\cite{danielewicz1}
\begin{equation}
\label{green_def}
G(1,2)=-i\frac{\Tr\{\hat{U}(t_0-i\beta,t_0)\mathcal{T_C}[\hat{\psi}_H(1)\hat{\psi}_H^{\dagger}(2)]\}}{\Tr\{\hat{U}(t_0-i\beta,t_0)\}},
\end{equation}
where $\mathcal{T_C}$ denotes the time ordering operator along the Keldysh contour $c$ (see Fig.\ref{fig:contour}) and $\hat{U}$ is
the time evolution operator. 
The average is taken over the grand canonical ensemble. We use the compact notation $1=(\mathbf{x}_1,t_1)$ and 
$2=(\mathbf{x}_2,t_2)$ where $\mathbf{x}=(\mathbf{r},\sigma)$ is a collective space-spin index.  
From the Green function it is possible to calculate any one particle property of the system.
For example the time-dependent density is given as
\begin{equation}
 \langle \hat{n}(1)\rangle =-iG^<(1,1^+),
\end{equation}
where  $t^+$  approaches $t$ from an infinitesimally later time $t^+ = t+\delta$. The equation of motion for the 
full system can be derived from the definition of the Green function Eq.~\eref{green_def} and reads
\begin{equation}
\label{EOM}
[i\partial_{t_1}-H(1)]G(1,2)=\delta(1,2)+\int_{c}\;\textrm{d}3\;\Sigma^{MB}[G](1,3) G(3,2),
\end{equation}
where $H$ is the one-body part of the Hamiltonian. The self-energy $\Sigma[G]$ incorporates 
all of the effects of exchange and correlation
in the central region and is a functional of the Green function~\cite{danielewicz1, baym_book}. 
In a localized basis the one-body part of the Hamiltonian~\eref{eq:hamiltonian}
and the many-body self-energy  can  be written in a block-matrix form 
\begin{eqnarray}
\label{eq:sigma}
&H=
\pmatrix{
 H_{LL} & H_{LC} & 0               \cr
 H_{CL} & H_{CC}  &H_{CR} 	\cr
 0                    & H_{RC}  &H_{RR} \cr
},
&
\;\;\;
\Sigma^{MB}=
\pmatrix{
 0  &       0      &  0  \cr
 0  &  \Sigma_{CC}^{MB}[G_{CC}] &  0  \cr
 0  &    0         &  0  \cr
},
\end{eqnarray}
where the $H_{\alpha \alpha}$ and $H_{CC}$ components describe the leads 
and the central system respectively, whereas the off-diagonal components describe the hopping 
between them~\cite{Petri}. We only consider the central region as interacting 
whereas the leads are effictively noninteracting. As a consequence, the many-body self-energy in Eq. \eref{eq:sigma} 
has non-vanishing elements only for the central region because the diagrammatic expansion starts and ends
with an interaction line. In this work the electronic interactions
are incorporated in  $\Sigma^{MB}[G]$ at HF, 2B and GW level~\cite{Petri}.

Solving the problem of an open infinite  system is equivalent to solving the problem of a closed system 
with an equation of motion which considers the leads through an embedding term~\cite{Petri}. In our case 
this reads (in the reminder of  this paper we will suppress the spatial indices of the objects involved)
\begin{equation}
\label{central_EOM}
\left[i\partial_t-H_{CC}(t)\right]G_{CC}(t,t') =\delta(t,t')+\int_{c}\,\textrm{d}\bar{t}\,
\left\{\left[\Sigma_{em}(t,\bar{t})+\Sigma_{CC}^{MB}[G_{CC}](t,\bar{t})\right]G_{CC}(\bar{t},t')\right\},
\end{equation}
where the embedding self-energy $\Sigma_{em}(t,t')$  accounts for the tunneling of electrons between 
leads and central region. In its general form, the embedding self-energy reads~\cite{Petri, petri_epl_08, Petri3}
\begin{equation}
\Sigma_{em}(t,t')=\sum_{\alpha} \Sigma_{em,\alpha}(t,t')
=\sum_{\alpha}H_{C\alpha}g_{\alpha\alpha}(t,t')H_{\alpha C}, 
\label{emb_self}
\end{equation}
where $g_{\alpha\alpha}(t,t')$ is the lead Green function and $H_{C\alpha},H_{\alpha C}$ is the 
coupling Hamiltonian between the central site and the leads.

The current through the lead $\alpha$ can be expressed in terms of Keldysh Green functions 
as \cite{Petri, petri_epl_08}
\begin{eqnarray}
\label{eq:current}
  \textrm{I}_{\alpha}(t)&=&2\textrm{Re}\bigg\lbrace \Tr_{C}\Big[ \int_{t_0}^{t}d\bar{t}
 [G_{CC}^<(t,\bar{t})\Sigma_{em,\alpha}^A(\bar{t},t)
+\int_{t_0}^{t}d\bar{t}G_{CC}^R(t,\bar{t})\Sigma_{em,\alpha}^<(\bar{t},t)]\nonumber\\
 &-i&\int_0^{\beta}d\bar{\tau}G_{CC}^{\rceil}(t,\bar{\tau})\Sigma_{em,\alpha}^{\lceil}(\bar{\tau},t)\Big]\bigg\rbrace,
\end{eqnarray}
where we integrated on the Keldysh contour and where the superscripts $A,R,<$ refer to advanced/retarded/lesser 
component of Green function/self-energy and $\rceil, \lceil$ are the mixed components having one time 
argument on a imaginary axis and  the other on the  real axis \cite{robert_06,dvl.2007,Petri}.  The trace is taken over the central region. 
The  current accounts for the initial  many-body and embedding effects through the last term 
in  equation \eref{eq:current} which is an integral over the vertical track.
Equation~\eref{eq:current} is a generalization of the Meir-Wingreen formula~\cite{ Meir_PRL_92}.
We further define the nonequilibrium spectral function 
\begin{equation}
\label{eq:spectral}
A(T,\omega)=-\textrm{Im}\, \textrm{Tr} \int\frac{\textrm{d}\tau }{2\pi } e^{i\omega\tau}\big[ G^>-G^<\big](T+\frac{\tau}{2},T-\frac{\tau}{2}),
\end{equation}
where $\tau=t-t'$ is a relative time and $T=(t+t')/2$ is an average time coordinate.
In equilibrium, this function is independent of $T$ and  
has peaks below the Fermi level at the electron removal energies of the system, 
while above the Fermi level it has peaks at the electron addition
energies. 
If the time-dependent external field becomes constant after some switching time, then also
the spectral function becomes
independent of $T$ after some transient period 
and has peaks at the addition and removal energies of the biased system.

\subsection{Time-dependent density functional theory}

TDDFT provides an alternative framework to describe electron transport through 
interacting systems. In TDDFT \cite{TDDFT_book}  the time-dependent 
density of the interacting system is obtained through time-propagation of a Kohn-Sham
system in an effective local potential. While in MBPT the 
correlation level is determined by the choice of the many-body self-energy, 
in TDDFT the main approximation is the functional 
used for the effective potential. The difficulty of TDDFT, compared 
to MBPT is the current lack of sufficient accurate approximations 
to the time-dependent exchange-correlation potential. 
However, the computational effort  is much lower 
compared to a MBPT propagation. 

The problem brought forward by considering an open 
system can be resolved in a very similar manner as in MBPT,
with the help of an embedding self-energy. 
The equation of motion for the $k$-th single-particle 
orbital projected onto the central region, $\psi_{k,C}(t)$, reads
\begin{equation}
\left[i\partial_t-\textrm{H}_{CC}(t)\right]\psi_{k,C}(t) = 
\int_{0}^t \,\textrm{d}\bar{t}\, \Sigma_{em}^R(t,\bar{t}) \psi_{k,C}(\bar{t}) 
+ \sum_{\alpha} H_{C \alpha} g_{\alpha \alpha}^R(t,0) \psi_{k,\alpha}(0),
\label{TDDFT_EOM}
\end{equation}
where $\Sigma_{em}^R(t,\bar{t})$ is the retarded embedding self-energy 
(see Eq.~(\ref{emb_self})) and $g_{\alpha \alpha}^R$ is the retarded lead Green 
function. The time-dependent density in the central region 
is  obtained by 
\begin{equation}
n(t) = \sum_k^{occ} |\psi_{k,C}(t)|^2 ,
\end{equation}
where the summation is taken over all occupied orbitals in the time-dependent Slater 
determinant \cite{attila}. The technique to propagate Eq.~(\ref{TDDFT_EOM}) 
 is described in detail in Ref.~\cite{ksarg.2005}. In this work we 
 used this approach at a level, in which, for the system studied,
the local exchange potential is equal to half the Hartree
potential of the localized site. The results were found, as expected, 
to be identical to those obtained from the Kadanoff-Baym approach at HF level.

\subsection{Steady-state density}

We begin our study of the bistable regime by deriving an equation for the density
on the localized site. This quantity
is given by the lesser Green function at equal times
\begin{equation}
 \langle \hat{n}(t)\rangle = -iG^{<}(t,t^+).
\end{equation} 
Since we consider  the steady-state, we can assume that in the long-time limit the 
Green functions depend only on the relative time coordinate $t-t'$.
In that case the Green function can be Fourier transformed with respect to the
relative time variable and the expression for the steady-state becomes
 \begin{equation}
n =\int \frac{\textrm{d}\omega}{2\pi i}G^<(\omega).
\label{density_integral}
\end{equation}
 The Green function appearing in this expression satisfies
 the equation 
 \begin{equation}
\label{eq:lsr_omega}
 G^<(\omega)=G^R(\omega)\Sigma_{tot}^<(\omega)G^A(\omega),
\end{equation}
where $\Sigma_{tot}^<(\omega)=\Sigma_{em}^<(\omega)+\Sigma^{MB,<}_{CC}(\omega)$
and where $G^A=[G^R]^{^*}$. The retarded Green function has the expression
\begin{equation}
G^R(\omega)=\left[\omega- \varepsilon_0-\textrm{Re}[\Sigma_{tot}^R(\omega)]
-i\textrm{Im}[\Sigma_{tot}^R(\omega)]\right]^{-1}. 
\end{equation}
For the tight-binding leads, that we consider,  the retarded embedding self-energy is
given by 
\begin{eqnarray}
\label{em_omega}
 \Sigma_{em,\alpha}^{R}(\omega)=\Lambda_\alpha(\omega)-\frac{i}{2}\Gamma_\alpha(\omega)
=\frac{V_{1\alpha,0}V_{0,1\alpha}}{2b^2}\cases{ \omega_{\alpha}-\sqrt{\omega_{\alpha}^2-4b^2}   &$, \omega_{\alpha}>2|b|    $\\
                                                \omega_{\alpha}+ \sqrt{\omega_{\alpha}^2-4b^2}  &$, \omega_{\alpha}<-2|b|   $\\
                                                \omega_{\alpha}- i\sqrt{4b^2-\omega_{\alpha}^2} &$, |\omega_{\alpha}|<2|b|  $\\
}
\end{eqnarray}
where $\omega_{\alpha}=\omega-a+\mu-U_{\alpha}$, with the lead-on-site parameter $a$ and  the 
hopping parameter between the lead sites $b$. The chemical potential is denoted by $\mu$, 
the applied bias for the lead $\alpha$ by $U_{\alpha}$, and $V_{1\alpha,0},V_{0,1\alpha}$ are the left/right couplings between 
leads and the central site. 
The lesser 
component of the embedding self-energy can be expressed as $\Sigma_{em,\alpha}^<(\omega)=if_{\alpha}(\omega)\Gamma_{\alpha}(\omega)$
where $f_\alpha$ is the Fermi distribution of lead $\alpha$ and $\Gamma_\alpha$ is defined in Eq. (\ref{em_omega}).

If we integrate the left hand side of Eq.~(\ref{eq:lsr_omega}) over all frequencies then
according to Eq.~(\ref{density_integral}) we obtain an expression for the density per spin $n$ on the localized site
\begin{equation}
\label{density_mw}
   n=\int_{-\infty}^{\infty}\frac{\textrm{d}\omega}{2\pi}\frac{\Gamma_L(\omega)f_{L}(\omega)
  +\Gamma_R(\omega)f_{R}(\omega)-i\Sigma^{MB,<}_{CC}(\omega)}{(\omega-\textrm{Re}
  [\Sigma^{MB,R}_{CC}(\omega)]-\Lambda(\omega))^2+(\textrm{Im}[\Sigma^{MB,R}_{CC}(\omega)
  ]-\Gamma(\omega)/2)^2},
\end{equation}
where $\Lambda=\Lambda_R+\Lambda_L$ and $\Gamma=\Gamma_R+\Gamma_L$.
This is a Meir-Wingreen-type equation for the density~\cite{Jauho_PRB_94, Meir_PRL_92} 
and is valid for transport through interacting systems.
Within the HF approximation for the IRLM one has 
$\Sigma^{MB,<}= \textrm{Im} [\Sigma^{MB,R}]=0$ and $\textrm{Re} [\Sigma^{MB,R}]=\mathcal{U} n$.
Note that we include the time-singular part of the self-energy in the definition of the retarded/advanced component, see {\em e.g.}, \cite{dvl.2007}.
Within this approximation, Eq. \eref{density_mw} now becomes a self-consistent fixed-point equation for the density $n$, since the value of the integral \eref{density_mw} depends on the density via the term $\mathcal{U} \, n$ in the denominator. 

\section{Results and discussion}

\begin{figure}[htbp]
  \centering
  \subfloat[]{\label{fig:spectrum_HF}\includegraphics[width=0.49\textwidth]{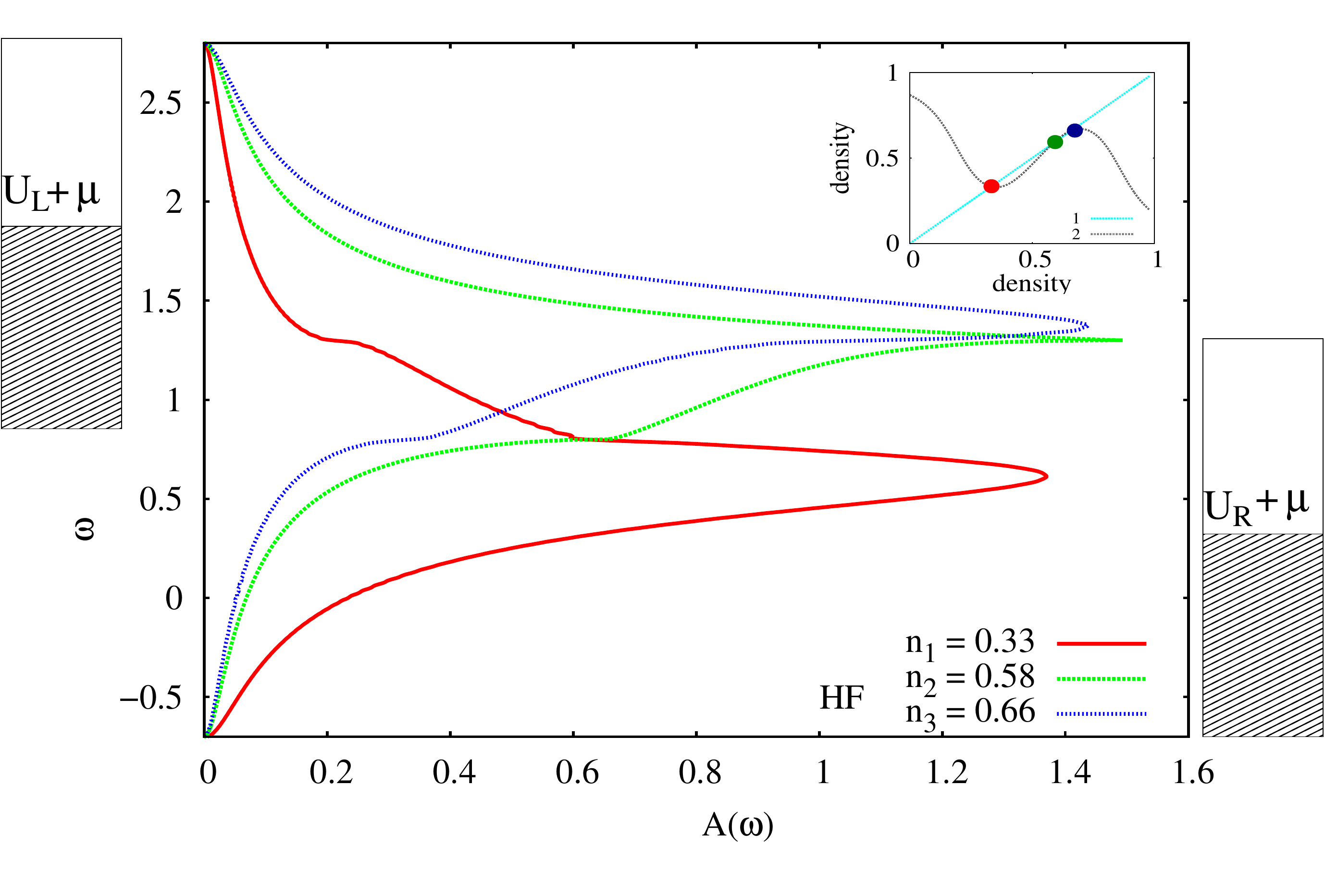}}
  \subfloat[]{\label{fig:spectrum_2BGW}\includegraphics[width=0.49\textwidth]{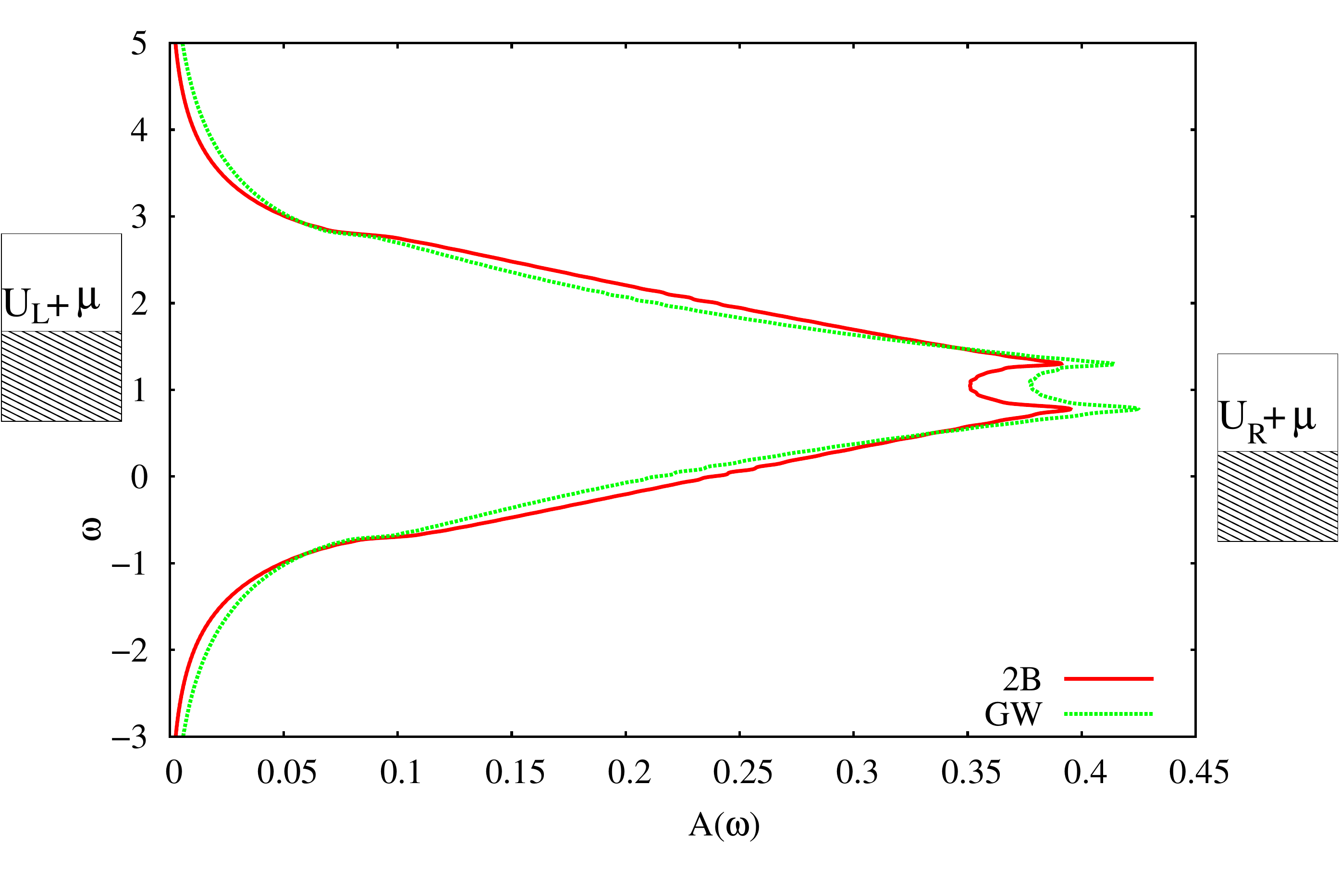}}      
  \caption{ a) Spectral functions for the HF approximation corresponding 
  to different steady-state solutions. {\em Inset}: The graphical solution of the integral in 
  Eq. \eref{density_mw}: (1) is the left hand side of the equation whereas  (2) is the value of the integral on the right hand side with given density $n$.  b) Steady-state spectral functions for the 2nd Born and GW approximations.}      
\end{figure}

We consider a biased system with the following parameters: 
$V_{0,1R}=V_{0,1L}=V=-0.35$, U$_{\textsc{l}}$ = 1.5, 
U$_{\textsc{r}}$ = 0.0, $\mathcal{U}$ = 2.0, $b$ = -0.5, $a = \mu$ and $\mu$ = 0.3, 
$\beta$ = 90.  The leads are half-filled such that the Fermi level 
of lead $\alpha$ is positioned at $\mu+U_{\alpha}$ and the width of the 
lead band is $[\mu+U_{\alpha}-2b,\mu+U_{\alpha}+2b]$.
With these parameters, within the HF approximation, the Meir-Wingreen 
approach \cite{Jauho_PRB_94, Meir_PRL_92} predicts 
three solutions for the steady-state density $n$ in Eq. (\ref{density_mw}). 
The three fixed points are shown in the inset of Fig.~\ref{fig:spectrum_HF},
where we display the left and right hand side of Eq.~(\ref{density_mw}).
The corresponding densities are $n_1=0.33$, $n_2=0.58$ and $n_3=0.66$, which should be
compared to the density of $n_0=0.28$ of the unbiased equilibrium state.

In  \ref{fig:spectrum_HF} we show the steady-state spectral functions corresponding to these three solutions,
which can be obtained directly from the retarded Green function $G^R (\omega)$.
The peak of the spectral function corresponding to 
density $n_1= 0.33$  is  positioned at the energy 0.5. 
We thus see that this spectral function is located in an energy range within the right lead energy-band
(see right side of Fig. \ref{fig:spectrum_HF}). 
The spectral function corresponding to density $n_3 = 0.66$ 
is peaked approximately at energy 1.5 and is located in an energy range within the left lead energy-band. 
The HF spectral function corresponding second solution with density $(n_2=0.58)$ is peaked on the top edge of the right 
lead energy-band and located in between the spectral functions corresponding to the densities $n_1$ and $n_3$.
By time-propagation (see below), we find  that the two solutions corresponding to densities $n_1$ and $n_3$ lead to 
stable steady-states, {\em i.e}, states that are reachable by time-propagation after applying a bias. On the other hand, the state corresponding to density $n_2$ is unstable and cannot be reached by time-propagation.
The spectral function corresponding to the state with density $n_2$ has a large overlap with the one of density 
$n_3$. This indicates that  (for sufficiently slow gate switching, see below) during the time-propagation more charge will flow onto the central site
resulting in a stable steady-state with density $n_3$.
From the analysis of the spectral functions we thus conclude that the density bistability 
in this system occurs when the spectral functions of the two stable solutions are localized well within one of the 
lead energy-bands and are well-separated. This happens exactly when
the leads have a small overlap and the system is  within the region of  negative differential resistance (NDR). 

Let us now consider the situation when we go beyond the HF approximation.
In Fig. \ref{fig:spectrum_2BGW} we show the steady-state spectral functions for the 2B  and GW approximations, as obtained
from time-propagation of the Kadanoff-Baym equations. With these approximations we found  only one steady-state solution, with very broad spectral function due to enhanced quasi-particle scattering at finite bias \cite{Petri}. The spectral
weight of the 2B and GW states is spread almost uniformly over the energy range from the bottom of the right
lead energy band to the top of left lead energy band and extends well outside the lead bands. 
We also observe two small peaks in these spectral functions which occur approximately at
 $0.8$ and $1.3$, corresponding to the edges of the lead energy bands.

\begin{figure}[htbp]
  \centering
  \subfloat[]{\label{fig:density_HF}\includegraphics[width=0.49\textwidth]{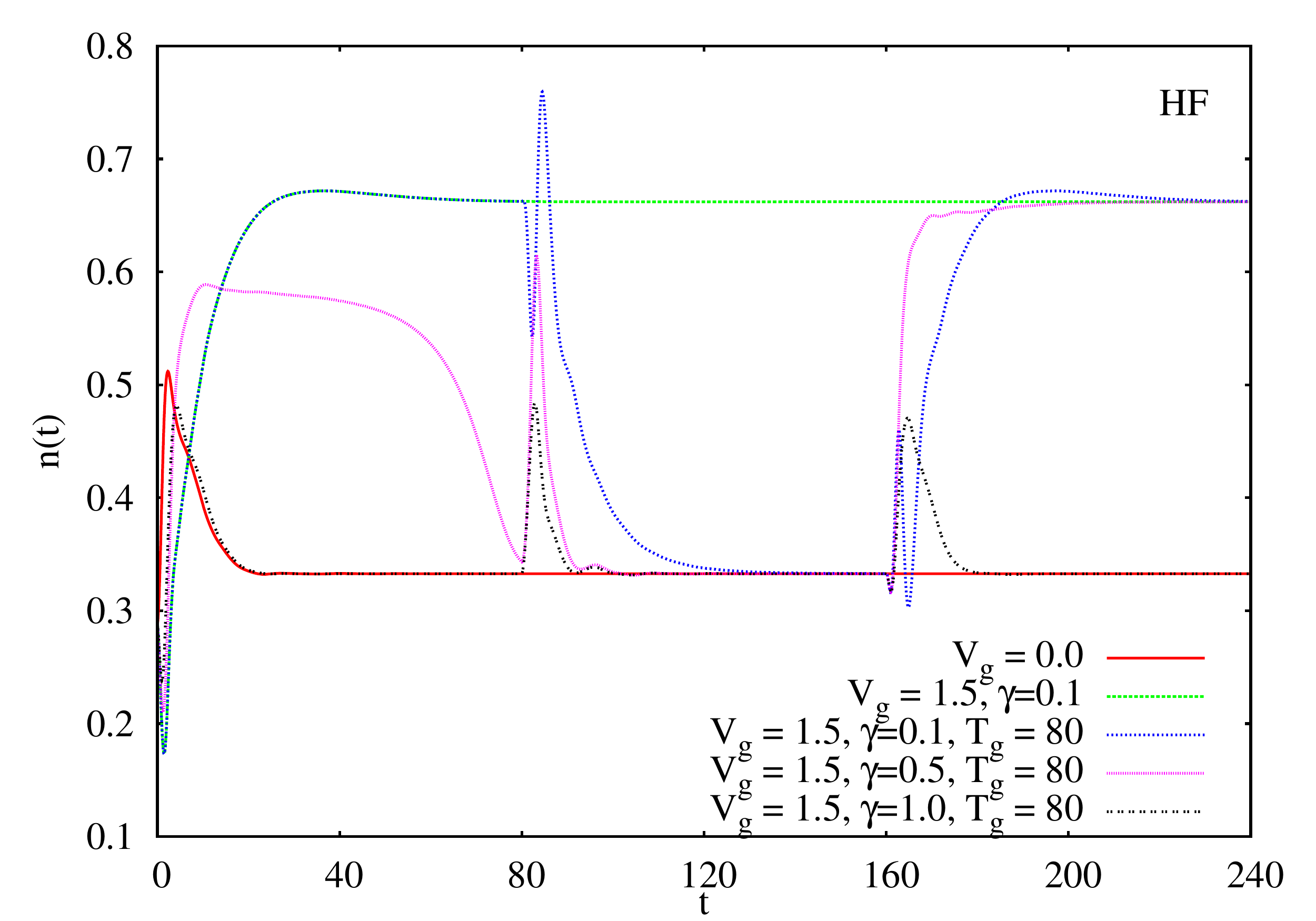}}
  \subfloat[]{\label{fig:current_HF}\includegraphics[width=0.49\textwidth]{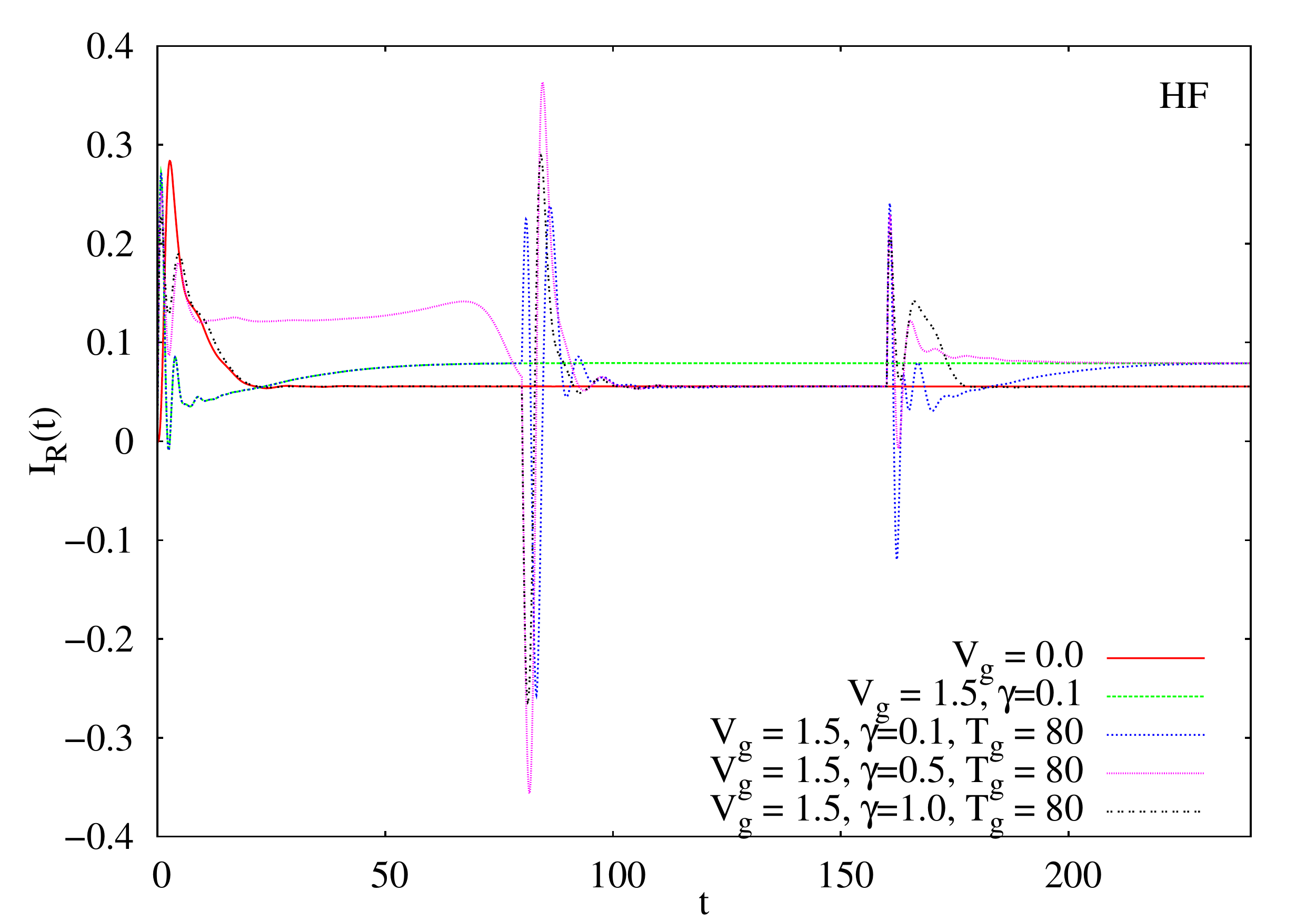}}      
  \caption{ a) Two different steady-state densities and the switching between the solutions by applying a gate 
  voltage in the form of the equation \eref{eq:gate}. b) Two different steady-state currents through right lead and the 
  switch between the solutions with a gate voltage of the form \eref{eq:gate}.}          
\end{figure}

We now go to the time domain and consider how the steady-states at HF level, corresponding to densities
$n_1 = 0.33$ and $n_3 = 0.66$, can be reached 
by time-propagation starting from the initially unbiased equilibrium state.
We generate these steady-states by applying time-dependent gate pulses
and we use the same technique to switch between them.
In this work we have used an exponentially decaying gate voltage of the form
\begin{equation}
\label{eq:gate}
 V_g(t)=\cases{
        V_ge^{-\gamma t}         & $,\;\textrm{if}\;\;0<t<T_{g}     $\\
        -V_ge^{-\gamma(t-T_{g})} & $,\;\textrm{if}\;\;T_{g}<t<2T_{g}$\\
        V_ge^{-\gamma(t-2T_{g}) }& $,\;\textrm{if}\;\;t>2T_{g}      $\\
        }.
\end{equation}  
The steady-state with density $n_1$ is obtained by time-propagation after applying a sudden constant bias $U_L(t)=U_L \theta (t)$ in the left lead, without applying the gate voltage.  This is displayed in Fig. \ref{fig:density_HF}.
The steady-state of highest density $n_3$ is obtained (in addition to switching on the sudden bias in the leads)
by applying an exponentially decaying gate voltage to the impurity site (with amplitude $V_g=1.5$, decay rate $\gamma=0.1$ and $T_g=\infty$).

In Fig. \ref{fig:density_HF} we show the time evolution of the density for
various switchings between the steady-states. After we apply the 
first (sufficiently slow) gate voltage of equation~\eref{eq:gate} we let the system reach the steady-state of density
$n_3$. At a time $T_g$ we apply a second gate voltage in the opposite direction. The system 
shows a transient behavior after which, the system reaches the steady-state of density 
$n_1$.  If we  apply the 
third gate voltage at time $2T_g$ (see equation \eref{eq:gate}), the system exhibits a short transient behavior and  attains again
the original steady-state of density $n_3$. 

Corresponding to the densities $n_1$ and $n_3$ there are two distinguishable solutions for the currents $I_R (t)$ flowing
into the right lead. These are 
shown in Fig. \ref{fig:current_HF} where the lower value of the current corresponds to the state of density $n_1$ and the 
higher value of the current corresponds to the state of density $n_3$. The various transients observed in the density correspond 
directly to the transients in current shown in Fig. \ref{fig:current_HF}. The frequency of the
strongly damped oscillations of the transients (observed upon switching between the states
in Fig.\ref{fig:current_HF}) is approximately given by the gate voltage, 
which causes a temporary change of $V_g$ in the level position.
As one can see from the Fig.~\ref{fig:density_HF}, if the decay rate $\gamma >0.5$, {\it i.e.}, when the switching is too fast, 
the steady-state of density $n_3=0.66$ cannot be reached from the initial ground state of density $n_0=0.28$, because 
the system does not have enough time to acquire sufficient density.

\begin{figure}[htbp]
  \centering
  \subfloat[]{\label{fig:contour_A}\includegraphics[width=0.49\textwidth]{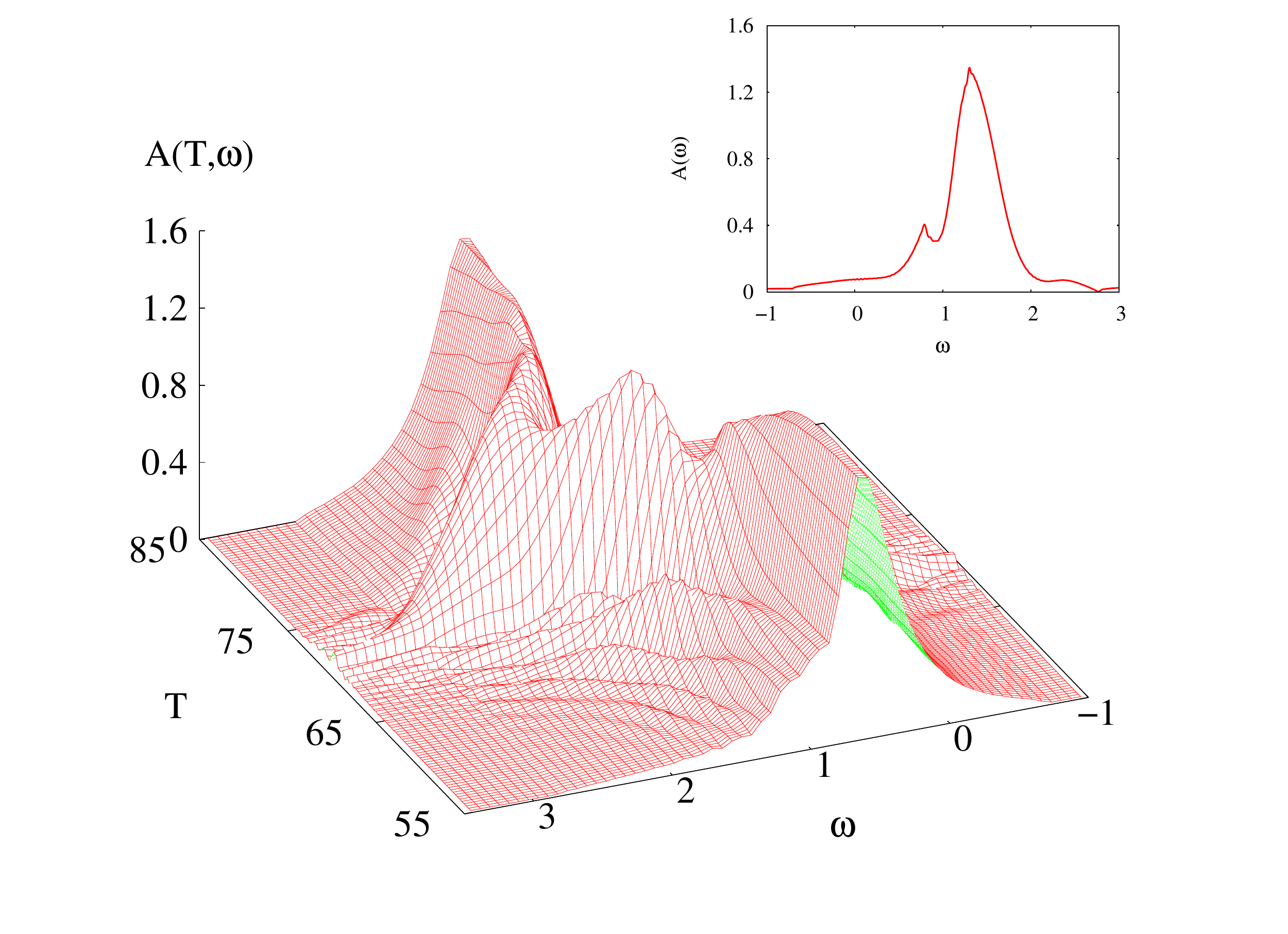}}                
  \subfloat[]{\label{fig:density_current_2BGW}\includegraphics[width=0.49\textwidth]{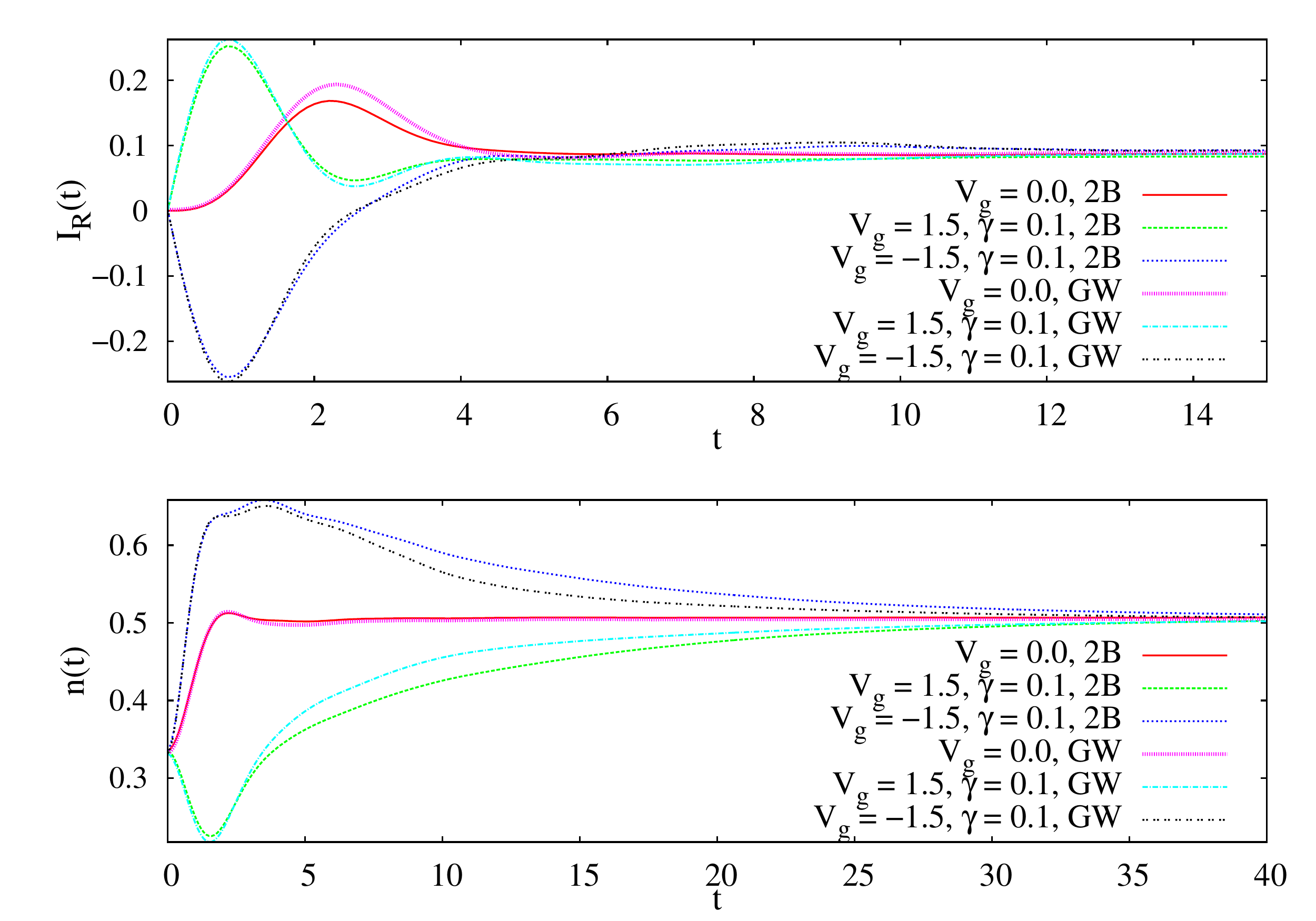}}
  \caption{ a) Nonequilibrium spectral function $A(T,\omega)$ within the HF-approximation for the switch from the density $n_1$ to density $n_3$. 
  {\em Inset}: Snapshot of spectral function corresponding to density $n_2$.
  b) The densities and currents calculated within the 2B and the 
  GW approximations with and without an exponentially decaying gate voltage.}
\end{figure}

In Fig. $\ref{fig:contour_A}$ we show, within the HF approximation, 
the nonequilibrium spectral function $A(T,\omega)$ of Eq.(\ref{eq:spectral}) for a
switch (with $T_g=60$, $V_g=1.5$, and $\gamma=0.1$) from the steady-state with density $n_1$ to the state with density $n_3$. 
When the gate is applied, the upper side of the spectral peak,
starts to oscillate. The spectral peak undergoes a transition and overshoots before it settles to a new value of $1.5$ within the right lead band. During this transition
we can observe the appearance of a sharp peak in the spectral function located approximately at energy $1.3$ 
  corresponding to the
top edge of the right lead band. This is shown in the inset of Fig.~\ref{fig:contour_A}, where we display a snapshot
of $A(T,\omega)$ at time $T=70$ at which the central site has density $n(T)=0.53$, which is very close to the density
of unstable state $n_2$. 
Therefore, although this state does not lead to a steady-state, it can still be observed in the nonequilibrium spectral
function.

In Fig. \ref{fig:density_current_2BGW}  we show the densities and the currents 
obtained within the 2B and the GW self-energy approximations. 
For these cases only one steady-state is obtained with a density of about $0.5$ on the central site.
We have applied a time-dependent gate voltage of the 
form $V_g(t)=V_ge^{-\gamma t}$ for $t >0$. If no gate is applied the transient regime is shorter 
and the density attains its steady-state value faster, compared to the case where the gate is applied. One 
can observe that the steady-state values of the currents and the densities for both GW and 2B are close to each other implying that 
for this system the single-bubble diagram, common to both approximations, plays a crucial role \cite{Petri}.

We conclude that for both the 2B  and the GW approximation the spectral functions are very broad
(see Fig.  \ref{fig:spectrum_2BGW}) which makes it impossible to locate these spectral functions within an energy range
in one of the lead energy-bands leading to two well-separated states. 
As a consequence the bistable regime is lost in the 2B and GW approximations, at least
for the parameters considered.

\section{Conclusions}

In this paper we have investigated the switching between the steady-states of an interacting resonant level model connected to 
 leads. For a given set of parameters we used 
a Meir-Wingreen type of approach to identify three different steady-state values for the density
within the HF approximation.  
We showed that by applying an exponentially decaying gate voltage pulse during the time-propagation, at 
Hartree-Fock level, we can reach two different solutions and switch between them.  
However, due to a strong quasi-particle broadening of 
the spectral function,  bistability is lost when the second Born and GW approximations are considered.
Hence treatment of many-body interactions 
beyond mean-field can destroy bistability and lead to qualitatively different results as compared to those at
mean-field level.

\nocite{fetter}
\section*{References}


\end{document}